\documentclass[12pt]{article}

\usepackage{hyperref}

\makeatletter \@addtoreset{equation}{section} \makeatother

\def\ads{AdS_{5}}
\def\be{\begin{equation}}
\def\ee{\end{equation}}
\def\ba{\begin{array}}
\def\ea{\end{array}}
\def\d{\partial}
\def\dps{\displaystyle}
\def\da{a_{\alpha}\frac{\partial}{\partial a_{\beta}}}
\def\db{b_{\alpha}\frac{\partial}{\partial b_{\beta}}}
\def\ba{\begin{array}}
\def\ea{\end{array}}
\def\d{\partial}
\def\dps{\displaystyle}
\def\da{a_{\alpha}\frac{\partial}{\partial a_{\beta}}}
\def\db{b_{\alpha}\frac{\partial}{\partial b_{\beta}}}

\newcommand{\half}{\frac{1}{2}}
\def\T{{\cal T}}
\def\sp{s_1+s_2-1}
\def\sm{s_1-s_2-1}

\begin{document}

\begin{flushright}
FIAN/TD/02/05
\end{flushright}

\vspace{0.51cm}

\begin{center}

{\Large\bf Mixed-symmetry massless gauge fields in $AdS_5$}

\vspace{.7cm}

\textsc{K.B.  Alkalaev}

\vspace{.7cm}

{\em I.E.Tamm Department of Theoretical Physics, P.N.Lebedev Physical
Institute,\\
Leninsky prospect 53, 119991, Moscow, Russia


}

\vspace{1cm}

\begin{abstract}

Free $AdS_5$ mixed-symmetry massless bosonic and fermionic gauge fields of
arbitrary spins are described by using $su(2,2)$ spinor language. Manifestly covariant action functionals  are
constructed and field equations are derived.

\end{abstract}

\end{center}


\vspace{1cm}

\section{Introduction}

    We continue the study of the five-dimensional higher-spin gauge theory \cite{SS1,VD5,A1,SS2,AV1}.
Our goal is to formulate manifestly covariant gauge invariant actions describing the free-field
dynamics of massless mixed-symmetry bosonic and fermionic gauge fields of arbitrary spins ${\bf
s}=$($s_1, s_2$). Our treatment of $\ads$ mixed-symmetry gauge fields is based on the frame-like
approach proposed in \cite{ASV,ASV2}. But  instead of the (spinor)-tensor language used in
\cite{ASV,ASV2}, we use a spinorial description of $\ads$ higher-spin fields based on the
well-known fact that the $\ads$ algebra $o(4,2)$ is isomorphic to $su(2,2)$. Therefore, $o(4,2)$
(spinor)-tensor fields can be described equivalently  as $su(2,2)$ multispinors. The main advantage
of the spinorial description is that bosonic and fermionic fields of any spin ${\bf s}=$($s_1,
s_2$) can be considered uniformly.

Although  the  $d$-dimensional analysis in  \cite{ASV,ASV2} certainly includes the $\ads$ case,
reformulating  these results in the spinor language  may be interesting in several aspects. In
general, such a reformulation is motivated by the desire to make a step towards a supersymmetric
nonlinear higher-spin gauge theory, which is of interest in the context of higher-spin version of
the $\ads/CFT_4$ correspondence (see \cite{Sezgin:2002rt,Bianchi:2004xi} for a review). The
nonlinear equations of motion for $AdS_d$ totally symmetric  bosonic fields and the  underlying higher-spin gauge algebras are now constructed \cite{Vasiliev:2003ev,Vasiliev:2004cm}, but
extending  them  to general mixed-symmetry fields remains the open problem. On the other hand, in the case
of $\ads$ higher-spin dynamics, there is a real possibility to construct nonlinear theory of
higher-spin fields of all symmetry types. Indeed, in five dimensions, one benefits  from
using the  isomorphism $o(4,2)\sim su(2,2)$. In particular, in the  spinor language,
(supersymmetric) higher-spin algebras were identified as certain star-product algebras with
$su(2,2)$ spinor generating elements \cite{FL,Vc}. There  also exist  manifestly covariant
formulations of free $\ads$ higher- spin dynamics\footnote{The light-cone equations of motion for
$AdS_d$  massless fields of arbitrary spins were constructed in
\cite{Metsaev,Metsaev:1998xg,Metsaev:1997nj} and the light-cone actions for generic $\ads$
mixed-symmetry massless fields were considered in \cite{Metsaev_d5}. Also, there are various
manifestly covariant Lagrangian formulations for particular examples of $AdS_d$ mixed-symmetry
gauge fields \cite{BMV,siegel,Zinoviev,deMedeiros:2003px}.} \cite{SS1,VD5,A1,SS2} and ${\cal
N}=0,1$ (supersymmetric) action functionals that  describe cubic interactions of  totally
symmetric $\ads$ fields \cite{VD5,AV1}.

The paper is organized as follows. In  Sec. 2, we describe the background
$\ads$ geometry in the spinor notations \cite{VD5}. In Sec. 3, we consider $\ads$
higher-spin gauge fields of mixed-symmetry type and
describe their multispinor and (spinor)-tensor forms.
In Sec. 4, we introduce  higher-rank tensors as functions of auxiliary spinor
variables \cite{VD5} and construct gauge transformations and linearized higher-spin curvatures.
In Sec. 5, we construct manifestly gauge invariant higher-spin actions. In Sec. 6, we derive
equations of motions and discuss  constraints for extra  fields.
We make concluding remarks in Sec. 7.

\section{$\ads$ background geometry in the spinor notations}

  Gravitational fields in $\ads$
are identified with 1-form connection that takes values in the $\ads$ algebra $su(2,2)$
\be
\label{grav}
\label{gf} \Omega(x) = dx^{\underline{n}}\:\Omega_{\underline{n}}{}^\alpha{}_\beta\:t_\alpha{}^\beta\;,
\ee
where $t_\alpha{}^\beta$ are basis elements of $su(2,2)$ and $\Omega^\alpha{}_\alpha = 0$ \footnote{ Throughout the paper we work within the mostly
minus signature and use the notations $\alpha, \beta, \gamma = 1\div
4$ for $su(2,2)$ spinor indices,
$\underline{m},\underline{n} = 0\div 4\;$ for  world indices, $a,b,c
= 0\div 4$ for tangent Lorentz $so(4,1)$ vector indices and $A,B,C =
0 \div 5$ for tangent $so(4,2)$ vector indices. We also use
condensed notations for a set of symmetric spinor indices $\alpha
(k)\equiv \alpha_1 \ldots \alpha_k$. Indices denoted by the same
letter are assumed to be symmetrized as $X^\alpha Y^\alpha \equiv
X^{\alpha_1}Y^{\alpha_2}+ X^{\alpha_2}Y^{\alpha_1}$. }.
The $su(2,2)$ gauge field in (\ref{grav}) decomposes into a frame field
and a Lorentz spin connection. This splitting can be  performed in a manifestly
$su(2,2)$ - covariant manner by introducing a compensator field \cite{compensator,VD5}, an
antisymmetric bispinor $V^{\alpha\beta}=-V^{\beta\alpha}$. The compensator is normalized
such that $V_{\alpha\gamma}V^{\beta\gamma}=\delta_\alpha{}^\beta$
and $V_{\alpha\beta}=\dps\half\varepsilon_{\alpha\beta\gamma\rho}V^{\gamma\rho}$.
The Lorentz subalgebra is identified with the stability algebra of the compensator.
This allows defining the frame field $E^{\alpha\beta}$ and Lorentz spin connection $\omega^{\alpha(2)}$
as \cite{VD5}
\be
\ba{l}
\dps E^{\alpha\beta} = DV^{\alpha\beta}\equiv dV^{\alpha\beta}
+\Omega^\alpha{}_\gamma V^{\gamma\beta} +\Omega^\beta{}_\gamma
V^{\alpha\gamma}\;,\;\;\;\; E^{\alpha\beta} V_{\alpha\beta}=0\;,
\\
\\
\dps \omega^\alpha{}_\beta = \Omega^\alpha{}_\beta
+\frac{\lambda}{2}E^{\alpha\gamma}V_{\gamma\beta}\;,
\\
\ea
\ee
where $\lambda$ is a cosmological parameter, $\lambda^2>0$. We note that because the compensator is Lorentz-invariant, we regard $V^{\alpha\beta}$ as a symplectic
metric that allows  raising and lowering spinor indices in a Lorentz-covariant way:
$X^\alpha=V^{\alpha\beta}X_\beta\;,\;Y_\alpha=Y^\beta V_{\beta\alpha}$.

The $AdS_5$ field strength corresponding to gauge field (\ref{gf}) has the form
\be
R_\alpha{}^\beta= d\:\Omega_\alpha{}^\beta
+\Omega_\alpha{}^\gamma\wedge\Omega_\gamma{}^\beta
\ee
and the background $\ads$ space is described by 1-form field
$\Omega_0^\alpha{}_\beta=(h^{\alpha\beta},\omega_0^{\alpha(2)})$, which satisfies
the zero-curvature condition \cite{V_obz2}
\be
\label{zcurv}
R_\alpha{}^\beta(\Omega_0) =0\;.
\ee

\section{Higher-spin fields}

To describe a spin-$(s_1,s_2)$ gauge field, we introduce
a pair of mutually conjugate
$su(2,2)$ traceless multispinors symmetric in the lower
and upper indices \cite{SS1,VD5,A1,SS2,ASV} \footnote{The complex conjugation operation is defined by the rule,
$\bar{X}_\alpha = X^\beta C_{\beta\alpha}\;, \; \bar{Y}^\alpha =C^{\alpha\beta}Y_\beta\;,$
where the bar denotes  complex conjugation and  $C^{\alpha\beta}=-C^{\beta\alpha}$ and
$C_{\alpha\beta}=-C_{\beta\alpha}$  are some real matrices such that $C_{\alpha\gamma}C^{\beta\gamma}=\delta_\alpha{}^\beta$ \cite{VD5}.},
\be
\label{multispinors}
\Omega^{\alpha(m)}{}_{\beta(n)}(x) \oplus \overline{\Omega}^{\beta(n)}{}_{\alpha(m)}(x)\,,
\qquad \Omega^{\alpha(m-1)\gamma}{}_{\beta(n-1)\rho}(x)\,\delta^\rho_\gamma=0\;,
\ee
which are 1-forms
\be
\label{1-form}
\Omega^{\alpha(m)}{}_{\beta(n)}(x) =
dx^{\underline{n}}\,\Omega_{\underline{n}}{}^{\alpha(m)}{}_{\beta(n)}(x)\;
\ee
with
\be
m=s_1+s_2-1\;,\quad n=s_1-s_2-1\;.
\ee
To decompose representations (\ref{multispinors}) of the $\ads$ algebra $su(2,2)$ into
representations of its Lorentz subalgebra, we  use of the compensator $V^{\alpha\beta}$. The result
of the reduction is given by
\be
\label{decomp}
\Omega^{\alpha(s_1+s_2-1)}{}_{\beta(s_1-s_2-1)}(x) =
\sum_{t=0}^{s_1-s_2-1}\omega^{\alpha(s_1+s_2-1)\gamma(t),\gamma(s_1-s_2-t-1)}(x)
V_{\beta(s_1-s_2-1),\gamma(s_1-s_2-1)}\;,
\ee
where the condensed notation
$V_{\alpha(k),\:\beta(k)}\equiv V_{\alpha_1\beta_1}V_{\alpha_2\beta_2}\ldots V_{\alpha_k\beta_k}$ is introduced.
The Lorentz algebra irreducible components
\be
\label{su_lor_fields}
\omega^{\alpha(s_1+s_2+t-1),\beta(s_1-s_2-t-1)}(x)\;,\quad  0\leq t\leq s_1-s_2-1
\ee
satisfy the Young symmetry condition
\be
\label{ic1}
\omega^{\alpha(s_1+s_2+t-1),\,\alpha\beta(s_1-s_2-t-1)}(x) =0\;
\ee
and contractions with $V_{\alpha\beta}$ are zero,
\be
\label{ic2}
\omega^{\alpha(s_1+s_2+t-1),\,\beta(s_1-s_2-t-1)}(x)V_{\alpha\beta} =0\;.
\ee
According to the analysis of \cite{SS1,VD5}, multispinors with $|m-n|=0$ correspond to totally
symmetric spin-$s_1$ bosonic fields and are self-conjugate. Other fields with $|m-n|\geq 1$ are described by a pair of mutually
conjugate multispinors and correspond either to totally symmetric fermionic spin-$s_1$ fields,
$|m-n|=1$ \cite{A1,SS2}, or to mixed-symmetry bosonic and fermionic fields, $|m-n|\geq 2$ \cite{VD5,SS2}.
We note that mixed-symmetry gauge fields necessarily occur
 in the spectrum of ${\cal N}\geq 2$ extended five-dimensional higher-spin gauge superalgebras, while
${\cal N}\leq 1$ (super)algebras describe totally symmetric fields \cite{Vc}.

To relate the spinor and (spinor)-tensor forms of mixed-symmetry field dynamics,
we examine the $o(4,1)$ (spinor)-tensor cousins of  multispinor fields (\ref{su_lor_fields})-(\ref{ic2}) at $s_2\neq 0$. The result is that
a collection  of $o(4,1)$ gauge fields is represented by complex-valued (spinor)-tensor
fields of the form
\be
\label{lorfields}
\omega^{a(s_1-1),\,b(s_2+t)}=dx^{\underline{n}}\,\omega_{\underline{n}}{}^{a(s_1-1),\,b(s_2+t)}\;,\quad
0\leq t \leq s_1 -s_2-1\;
\ee
for bosonic mixed-symmetry fields and
\be
\label{Florfields}
w^{\alpha\,|\, a(s_1-1),\,b(s_2+t)}
=dx^{\underline{n}}\,w_{\underline{n}}{}^{\alpha\,|\,a(s_1-1),\,b(s_2+t)}\;,
\quad
\alpha =1\div 4\;,
\quad 0\leq t \leq s_1 -s_2-1\;
\ee
for fermionic mixed-symmetry fields ($\alpha$ denotes a five-dimensional Dirac spinor index).
In both cases, fields  (\ref{lorfields}) and (\ref{Florfields}) have the Young symmetry property and are traceless (bosons) or gamma-transverse (fermions).

In accordance with the nomenclature in \cite{ASV}, fields  (\ref{lorfields}) and (\ref{Florfields}) with
the parameter $t\geq 1$ are called extra fields.  Fermionic field (\ref{Florfields}) at $t=0$ is called the
physical field. To classify the bosonic field  in (\ref{lorfields}) with $t=0$, we decompose it into real and
imaginary parts as
\be
\label{BcomplLor}
\omega^{a(s_1-1),\,b(s_2)}  = \omega_1{}^{a(s_1-1),\,b(s_2)}+i\,\omega_2{}^{a(s_1-1),\,b(s_2)}\;.
\ee
Using the five-dimensional Levi-Civita
symbol,  one of the fields, $\omega_1$ or $\omega_2$, can be dualized into a field with one index in
the  third row, for example,
\be
\label{BcomplLorDual}
\omega^{a(s_1-1),\,b(s_2)}  = \omega_1{}^{a(s_1-1),\,b(s_2)}
+i\,\epsilon^{abcde} \omega_2{}^{a(s_1-2)}{}_{c,\,}{}^{b(s_2-1)}{}_{d,\,e}\;\;,
\ee
where the  dual three-row field $\omega_2^{a(s_1-1),\,b(s_2),\,e}$ is traceless and has the
Young symmetry property. We call the real part
$\textrm{Re} \,\omega^{a(s_1-1),\,b(s_2)}$ of field (\ref{BcomplLorDual}) the physical field.
The imaginary part $\textrm{Im} \,\omega^{a(s_1-1),\,b(s_2)}$ of this field
 is called the auxiliary field.
In fact, any bosonic field (\ref{lorfields}) can be represented as a pair of real fields
with one of them having one index in a third row. The resulting collection of real Lorentz-covariant
fields is described by three-row $o(4,1)$ Young tableaux arising as a decomposition of
a certain $o(4,2)$ three-row Young tableau \cite{VD5,ASV}.

\section{Higher-spin linearized curvatures}

The analysis of linearized curvatures in this section is close to
the analysis in the previous papers on totally symmetric fields \cite{VD5,A1}.
It turns out that the general form of gauge transformations for
mixed-symmetry fields remains intact  except for an additional
dependence on the spin $s_2$ and the appearance of a nonzero operator ${\cal T}_0$
for bosonic nonsymmetric fields (see (\ref{T0}), (\ref{Delta})). This last feature is not typical for bosonic systems
and is a reflection of an implicit presence the Levi-Civita symbol in the definition of real
bosonic components of complex-valued fields (\ref{lorfields}).
The calculation of the bosonic operator ${\cal T}_0$ is the main result in this section.

We introduce auxiliary commuting variables $a_\alpha$ and $b^{\beta}$ transforming under the fundamental and
the conjugate fundamental representations of $su(2,2)$. It is convenient to represent higher-spin
fields (\ref{1-form}) as functions of auxiliary variables
\be
\label{gen} \Omega(a, b|x) =
\Omega^{\alpha(\sp)}{}_{\beta(\sm)}(x)\:a_{\alpha(\sp)}b^{\beta(\sm)}
\;,
\ee
where
\be
a_{\alpha(m)} = a_{\alpha_1}\cdots a_{\alpha_m}\;,
\qquad
b^{\beta(n)} = b^{\beta_1}\cdots b^{\beta_n}\;.
\ee
The corresponding five-dimensional linearized higher-spin curvature is given by
\be
\label{curv}
R(a,b|x) = d\:\Omega(a,b|x) +
\Omega_0{}^\alpha{}_\beta (b^\beta\frac{\d}{\d b^\alpha}-a_\alpha\frac{\d}{\d
a_\beta})\wedge \Omega(a,b|x)\;,
\ee
where is the background 1-form connection $\Omega_0{}^\alpha{}_\beta$ satisfies the
zero-curvature condition (\ref{zcurv}).
The linearized (Abelian) higher-spin transformation are
\be
\label{hstr}
\delta\Omega(a,b|x)=D_0\xi(a,b|x)\;,
\ee
where the background covariant derivative is given by

\be
D_0=d +\Omega_0{}^\alpha{}_\beta (b^\beta\frac{\d}{\d
b^\alpha}-a_\alpha\frac{\d}{\d a_\beta})\;.
\ee
Condition (\ref{zcurv}) implies that $\delta R(a,b|x)=0$. The Bianchi
identities have the form
\be
D_0R(a,b|x)=0\;.
\ee

In the subsequent analysis,  we  use two sets of the differential
operators  in auxiliary  variables  \cite{VD5},

\be
\label{S}
S^- = a_\alpha \frac{\d}{\d b^\beta}V^{\alpha\beta}\:,\qquad
S^+ = b^\alpha \frac{\d}{\d a_\beta} V_{\alpha\beta}\:,\qquad
\dps S^0 = N_b - N_a
\ee
and
\be
\label{T}
T^- = \frac{1}{4} \frac{\partial^2 }{ \partial a_\alpha
\partial b^\alpha }\:,\qquad T^+ = a_\alpha b^\alpha \:,\qquad
T^0 = \frac{1}{4}( N_a +N_b  +4 )\;,
\ee
where

\be
\label{N} N_a=a_\alpha\frac{\d}{\d a_\alpha} \qquad {\rm and}
\qquad N_b=b^\alpha\frac{\d}{\d b^\alpha}\;.
\ee
With (\ref{S}) and (\ref{T}),  the irreducibility conditions for $\Omega(a,b)$ are reformulated as
\be
T^- \Omega(a,b)=0\;,\qquad (S^0+2s_2)\Omega(a,b) =0\;.
\ee
As demonstrated in Sec. 3, the higher-spin gauge field
$\Omega$  decomposes into  Lorentz subalgebra representations in  accordance with formula (\ref{decomp}).
In terms of operators (\ref{S}) and (\ref{T}), formula (\ref{decomp}) is rewritten as
\be
\label{expan}
\Omega(a,b|x) = \sum_{t=0}^{\sm} (S^+)^t \:\omega^t(a,b|x),
\ee
where
\be
\omega^t(a,b|x) = \omega^{\alpha(s_1+s_2+t-1),\; \beta(s_1-s_2-t-1)}(x)\:
a_{\alpha(s_1+s_2+t-1)} b_{\beta(s_1-s_2-t-1)}
\ee
are Lorentz-covariant gauge fields (\ref{su_lor_fields}). The irreducibility conditions
in (\ref{ic1}) and (\ref{ic2}) become
\be
\label{irr}
S^-\omega^t(a,b) = 0\;, \qquad T^-\omega^t(a,b) = 0\;.
\ee

 Higher-spin gauge symmetry (\ref{hstr})
requires  the  bosonic and fermionic Lorentz-covariant higher-spin curvatures $r^t$
and gauge transformations to be given by
\be
\label{taus}
r^t = {\cal D} \omega^t  +{\cal T}^-\omega^{t+1} +\lambda\,{\cal T}^0\omega^t
+\lambda^2\,{\cal T}^+\omega^{t-1}\;,
\ee

\be
\label{taus2}
\delta\omega^t = {\cal D} \xi^t  +{\cal T}^-\xi^{t+1} +\lambda\,{\cal T}^0\xi^t
+\lambda^2\,{\cal T}^+\xi^{t-1}\;,
\ee
where 0-forms $\xi^t$ are Lorentz-covariant gauge parameters and  ${\cal D}$ is the
background Lorentz-covariant derivative
\be
\label{LD}
{\cal D} =d+w_0^{\alpha}{}_{\beta}(\da+\db)\;.
\ee
The operators ${\cal T}^-$, ${\cal T}^+$ and ${\cal T}^0$ have the form

\be \label{T+} {\cal T}^+
= (1-\frac{\Delta^2}{(S^0)^2})\;h^{\alpha}{}_{\beta}a_{\alpha}\frac{\partial}{\partial
b_{\beta}}\;,
\ee

\be
\label{T0}
{\cal T}^0=-\frac{\Delta}{S^0}\; h^{\alpha}{}_{\beta}(\db - \da
+\frac{2}{S^0-2}(b_\gamma\frac{\d}{\d a_\gamma})a_{\alpha}\frac{\partial}{\partial
b_{\beta}})\;,
\ee

\be
\label{T-}
\ba{c}
\dps{\cal T}^- = \frac{1}{1-S^0}\;h^{\alpha}{}_{\beta}((2-S^0)b_{\alpha}\frac{\partial}{\partial
a_{\beta}} + b_\gamma\frac{\d}{\d a_\gamma}(\db - \da)
\\
\\
\dps+
\frac{1}{S^0-3}\;(b_\gamma\frac{\d}{\d a_\gamma})^{2}a_{\alpha}\frac{\partial}{\partial
b_{\beta}})\;,

\ea
\ee
where the parameter $\Delta$ takes the values
\be
\label{Delta}
\Delta = \left\{ \ba{l}
\;\;2s_2\,,\;\;\;\; {\rm \;for\;\; bosons} \,, \\ 2s_2+1\,,\;{\rm for\;\; fermions}\,,
\ea \right.
\ee
and  satisfy the relations
\be
\ba{l}
\dps \{\T^0,\T^- \}=\{\T^0,\T^+\} = 0,
\\
\\
(\T^-)^2=0\;,\;(\T^+)^2=0\;,
\\
\\
{\cal D}^2+\lambda^2\, \{\T^-,\T^+\}+\lambda^2\,(\T^0)^2  =0\;.
\\

\ea
\ee
We note that the coefficients in (\ref{T+})-(\ref{T-}) can be changed by
field redefinitions of the form
$\widetilde{\omega}^t=C(t,s)\omega^t$ with $C\neq 0$.

\section {Higher-spin action}

Before the considering actions for arbitrary mixed-symmetry gauge fields, we
examine  the case of the simplest nonsymmetric bosonic field of spin $(2,1)$ described by
$1$-form $\Omega^{\alpha(2)}(x)$. Up to total derivative terms, the action functional  has the unique  form
\be
\label{huact}
{\cal S}_2^{(2,1)} = \int_{{\cal M}^5} h{}^{\alpha}{}_{\beta}\wedge R^{\beta\gamma}\wedge
\bar{R}_{\gamma\alpha}\;,
\ee
where $h{}^{\alpha}{}_{\beta}$ is the background $\ads$ frame field and the curvature is
\be
R^{\alpha(2)}= D_0\Omega^{\alpha(2)}\equiv
{\cal D}\Omega^{\alpha(2)}+\lambda \,h^\alpha{}_\gamma\wedge\Omega^{\gamma\alpha}\;.
\ee
The equations of motion resulting from the action (\ref{huact}) are
\be
\label{hueq}
H_2{}^{\alpha}{}_{\gamma}\wedge R^{\gamma\beta} +
H_2{}^{\beta}{}_{\gamma}\wedge R^{\gamma\alpha} = 0\;
\ee
plus the complex-conjugated equations for $\bar{\Omega}_{\alpha\beta}$. We note that these bosonic
equations are of 1-st order, which makes them similar to fermionic equations. But as discussed
in Sec. 3, the real and imaginary parts of the complex-valued field $\Omega^{\alpha(2)}(x)$
are regarded as physical and  auxiliary fields (\ref{BcomplLorDual}), with  the auxiliary
field being expressed by virtue of its equation of motion  in terms of  first derivatives of the physical field.
To describe  this mechanism in more detail, we consider
the tensor form of  action (\ref{huact}). According to (\ref{BcomplLor}), the $o(4,1)$ field
isomorphic to $\Omega^{\alpha(2)}(x)$ is
\be
\label{omega}
\omega^{[ab]}= \omega_1^{ab} +i \omega_2^{ab}\;.
\ee
The corresponding linearized curvature and gauge transformations have the forms
\be
\label{curv_gauge}
R^{ab}={\cal D}\omega^{ab}-\frac{i\lambda}{2}\epsilon^{abcde}\,h_c\wedge\omega_{de} \;,
\quad
\delta\omega^{ab} = {\cal D} \xi^{ab}-\frac{i\lambda}{2}\epsilon^{abcde}\,h_c \,\xi_{de} \;,
\ee
where ${\cal D}$ is the background Lorentz-covariant derivative, $\xi^{ab}$ is a 0-form complex gauge
parameter, and $h^a$ is the background frame field.
We note that the terms in (\ref{curv_gauge})  involving the Levi-Civita symbol are in fact the operator
${\cal T}^0$ expressed in the spinor notation by formula (\ref{T0}). The Bianchi identities are
\be
{\cal D} R^{ab}-\frac{i\lambda}{2}\epsilon^{abcde}\, h_c\wedge R_{de}=0\;.
\ee
The action has a form analogous to (\ref{huact})
\be
\label{huact2}
{\cal S}_2^{(2,1)} = \int_{{\cal M}^5} \epsilon_{abcde}h^e\wedge R^{ab}\wedge \bar{R}^{cd}\;,
\ee
where $\bar{R}^{cd}$ is complex-conjugate curvature (\ref{curv_gauge}).
The equations of motion are
\be
H_a{}^c\wedge R_{cb} - H_b{}^c\wedge R_{ca} = 0\;,\qquad H_{ab}\stackrel{\rm{def}}{=}h_a\wedge h_b\,
\ee
plus the complex-conjugate equations.

 To clarify the dynamical content of these equations, we regard  the real or imaginary part of the field $\omega^{ab}$ given by (\ref{omega}) as dualized auxiliary field, for example,
\be
\omega_1^{ab} = \omega_1^{ab}\;,\qquad \omega_2^{ab} = \frac{1}{\lambda}\,\epsilon^{abcde}\omega_{2\;cde}\;,
\ee
where $\omega_1^{ab}$ and $\omega_2^{abc}$ are the physical and  auxiliary fields with
antisymmetric indices and the factor $\lambda^{-1}$ is introduced to express the fact that mass
dimensions of the physical and auxiliary fields are different.
These fields can be unified into a single $o(4,2)$ field $\Omega^{[ABC]}$ \cite{ASV}.
It can be shown that the action (\ref{huact2}) can be rewritten  as
\be
\label{heact3}
{\cal S}_2^{(2,1)} = \frac{1}{\lambda^2}\int_{{\cal M}^d}\;\epsilon_{ABCDEF}\,h^{E}V^F\wedge R^{ABM}\wedge R^{CDN}\,V_M V_N\,
\ee
In this form, the action coincides with  the action for the  $AdS_d$ "hook" field explicitly studied in \cite{ASV}.
We note that the flat limit of  action (\ref{huact2}) (or, equivalently, (\ref{heact3}))
yields the dual description of the  spin-$2$ field, which precisely corresponds to  Curtright action \cite{curt}.
Another comment is that described procedure for  unifying  dynamical and auxiliary fields into a
single complex-valued field was used to study the so-called odd-dimensional
self-duality for massive antisymmetric tensor fields in Minkowski space \cite{gauged_sugra}.

\subsection{Action for nonsymmetric $\ads$ gauge fields }

In what follows, we construct free actions describing mixed-symmetry
bosonic and fermionic gauge  fields in $AdS_5$. The case of totally
symmetric fields was considered in \cite{VD5,A1}.

As in \cite{VD5,A1,ASV}, we seek mixed-symmetry field action functionals  in the form
\be
\label{act}
{\cal S}^{(s_1, s_2)}_2 =\int_{{\cal M}^5} \hat{H}\wedge  R^{s_1,s_2}(a_1,b_1)\wedge
\bar{R}{}^{s_1,s_2}(a_2,b_2)|_{a_i=b_i=0}\;,
\ee
where $R^{s_1,s_2}$ is linearized higher-spin curvature (\ref{curv})
and $\hat{H}$ is the  1-form differential operator
\be
\label{H}
\ba{c}
\dps \hat{H}=
\Big(\alpha(p,q) h_{\alpha\beta} \frac{\d^2}{\d a_{1\alpha} \d
a_{2\beta}}\hat{b}_{12}
+\beta(p,q) h^{\alpha\beta} \frac{\d^2}{\d b_1^\alpha \d
b_2^\beta}\hat{a}_{12}
\\
\\
\dps
+\gamma(p,q) h_\alpha{}^\beta \frac{\d^2}{\d a_{2\alpha} \d b_1^\beta}\hat{c}_{12}
+\zeta(p,q) h_\alpha{}^\beta \frac{\d^2}{\d a_{1\alpha} \d b_2^\beta}\hat{c}_{21}
\Big)\,(\hat{c}_{12})^{2s_2}\;.
\ea
\ee
Here $h_{\alpha}{}^{\beta}$ is the background frame field and the
coefficients $\alpha,\;\beta,\;\gamma$ and $\zeta$  are functions of operators
\be
p=\hat{a}_{12}\hat{b}_{12}\;,\qquad q=\hat{c}_{12}\hat{c}_{21}\;,
\ee
where
\be
\label{abg}
\ba{cc}
\dps\hat{a}_{12} = V_{\alpha\beta}\frac{\d^2}{\d a_{1\alpha} \d
a_{2\beta}}\;,&
\qquad\dps \hat{b}_{12} = V^{\alpha\beta}\frac{\d^2}{\d b_1^\alpha \d
b_2^\beta}\;,
\\
\\
\dps\hat{c}_{12} = \frac{\d^2}{\d a_{1\alpha} \d b_2^\alpha}\;,&
\dps\hat{c}_{21} = \frac{\d^2}{\d a_{2\alpha} \d b_1^\alpha}\;.
\\
\\
\ea
\ee
These functions are responsible for various types of index  contractions
between the frame field and curvatures. The action is invariant under complex
conjugation $\bar{{\cal S}}_2={\cal S}_2$ when the  coefficients
$\alpha,\;\beta,\;\gamma$ and $\zeta$ are real.

Because the general variation of the linearized curvatures is $\delta R= D_0\delta\Omega$
and because the action is formulated in an $AdS_5$ covariant way,
integrating by parts yields the variation
\be
\label{var} \delta{\cal S}^{(s_1, s_2)}_2 = \int_{{\cal M}^5} D_0\hat{H}\wedge
\delta\Omega(a_1,b_1)\wedge \bar{R}(a_2,b_2))|_{a_i=b_i=0}\;\;\;
+\;\;\;{\rm c.c.\;\; part}\;.
\ee
The derivative $D_0$ produces the frame field each time it hits
the compensator $D_0 V^{\alpha\beta}=h^{\alpha\beta}$. Taking $D_0 h^{\alpha\beta}
=0,\;h_\alpha{}^{\beta} =h_{\alpha\gamma}V^{\beta\gamma}$ into account  and
using  the notation  $H^{\alpha\beta}=H^{\beta\alpha} =
h^\alpha{}_\gamma\wedge h^{\beta\gamma}$, we find
\be
\label{varH}
\ba{c}
D_0\hat{H}=
\dps \Big(\rho_1 H_\alpha{}^\beta \frac{\d^2}{\d a_{2\alpha} \d b_1{}^\beta }\hat{c}_{12}
+\rho_2 H_\alpha{}^\beta \frac{\d^2}{\d a_{1\alpha} \d b_2{}^\beta}\hat{c}_{21}
\\
\\
\dps +\rho_3 H_{\alpha\beta} \frac{\d^2}{\d a_{1\alpha} \d a_{2\beta}}
\hat{b}_{12}
 +\rho_3 H^{\alpha\beta} \frac{\d^2}{\d b_1^\alpha \d b_2^\beta}
\hat{a}_{12}\Big)(\hat{c}_{12})^{2s_2}\;,
\\

\ea
\ee
where
\be
\label{rhos}
\ba{l}
\dps\rho_1 =\half\,\Big(1+p\frac{\d}{\d p}\Big)\Big(-2\gamma(p,q) + (\alpha+\beta)(p,q)\,\Big)\,,
\\
\\
\dps\rho_2 = \half\,\Big(1+p\frac{\d}{\d p}\Big)\Big(-2\zeta(p,q) + (\alpha+\beta)(p,q)\,\Big)\,,
\\
\\
\dps\rho_3= \half\, q\frac{\d }{\d p}\Big(\zeta(p,q)-\gamma(p,q)\,\Big)\,.

\ea
\ee
For the trivial solution $\rho_i=0$, the covariant derivative of
$\hat{H}$ vanishes, $D_0 \hat{H}=0$, and the corresponding action
functional is a total derivative. It follows from (\ref{rhos}) that
$\rho_i=0$ whenever
\be
\label{top}
(\alpha+\beta)(p,q) = 2\gamma(p,q)\;,
\qquad
\zeta(p,q) = \gamma(p,q)\;.
\ee
Clearly, by adding total derivatives with the coefficients satisfying
(\ref{top}), we can always set $\gamma = 0$ and $\beta=0$ in the action (\ref{act}), (\ref{H}).

Generally,  action (\ref{act}) does not describe massless higher-spin fields, because there are too many nonphysical dynamical variables associated with the
extra fields. To eliminate the corresponding degrees of
freedom, we must fix the operator $\hat{H}$ in an appropriate form by virtue
of \textit{the decoupling condition}  \cite{LV,vf,ASV}. It requires
the variation of the quadratic action with respect to the extra fields is identically zero,
\be
\label{exdc} \frac{\delta {\cal S}_2^{(s_1, s_2)}}{\delta\omega^{t>0}} \equiv 0\;.
\ee
To analyze the extra field decoupling condition, we observe that all gauge fields of the extra type
can be combined into a single irreducible $su(2,2)$ tensor $\xi(a,b)$  satisfying $(N_a-N_b-2s_2-2)\xi(a,b)=0$.
Then, the  variation  of the extra fields becomes
\be
\label{ex2}
\delta\Omega^{extra}(a,b)=S^+\xi(a,b)\;
\ee
and the extra field decoupling condition (\ref{exdc}) amounts to
\be
\label{eq2}
\ba{l}
\dps \Big(\frac{\d }{\d p} - \frac{\d}{\d q}\Big)  (q\rho_2)+\rho_3 =0 ,
\\
\\
\dps\Big(\frac{\d }{\d p} - \frac{\d }{\d q}\Big)\rho_3 =0 ,
\\
\\
\dps\rho_1 +\rho_3 =0.

\ea
\ee
Modulo total derivative contributions (\ref{top}), the general solution to the system (\ref{eq2})
is
\be
\ba{c}
\gamma(p,q)=0\,,
\quad
\beta(p,q)=0\,,
\quad
\dps \zeta(p,q)=\zeta^{(0)} \frac{(p+q)}{q}^{s_1-s_2-1}\,,
\\
\\
\dps \alpha(p,q) = -\zeta^{(0)}(s_1-s_2-1)\int_0^1 d\tau(p\tau+q)^{s_1-s_2-2} =
\\
\\
\dps
=-\zeta^{(0)}\sum_{k=0}^{s_1-s_2-2}\frac{(s_1-s_2-1)!}{(k+1)!(s_1-s_2-k-2)!} p^k\,q^{s_1-s_2-k-2}
\;.
\ea
\ee
The factor $q^{-1}$ appearing in $\zeta(p,q)$ can be removed by
redefining $\zeta(p,q)\rightarrow q\zeta(p,q)$. This operation
does remove the singularity because the last term in the operator
$\hat{H}$ in (\ref{H}) contains the combination $\hat{c}_{21}(\hat{c}_{12})^{2s_2}$,
which is always
$q(\hat{c}_{12})^{2s_2-1}$  by the definition of  $q$ in (\ref{abg}). An overall factor $\zeta^{(0)}$ in front
of the action of a given spin $(s_1,s_2)$ cannot be fixed from the
analysis of the free action and represents the residual ambiguity in
the coefficients.

\section{Equations of motion and constraints}

To obtain equations of motion, we rewrite the nontrivial part of  variation (\ref{var}) as
\be
\label{var_1}
\ba{c}
\dps
\delta{\cal S}^{(s_1, s_2)}_2 = -\frac{\zeta^{(0)}(s_1-s_2-1)}{2}\int (p+q)^{s_1-s_2-2}
\Big(\frac{(s_1-s_2+1)q+2(s_1-s_2)p}{s_1-s_2-1}\,H_\alpha{}^\beta \frac{\d^2}{\d a_{1\alpha} \d b_2{}^\beta}
\\
\\\dps + H_\alpha{}^\beta \frac{\d^2}{\d a_{2\alpha} \d b_1{}^\beta}(\hat{c}_{12})^2
-H_{\alpha\beta} \frac{\d^2}{\d a_{1\alpha} \d a_{2\beta}}\hat{b}_{12}\hat{c}_{12}
\dps -H^{\alpha\beta} \frac{\d^2}{\d b_1^\alpha \d b_2^\beta}\hat{a}_{12}\hat{c}_{12}\Big)(\hat{c}_{12})^{2s_2-1}
\\
\\
\wedge r^0(a_1,b_1)\wedge \delta \bar{\omega}^0(a_2,b_2) + \;\;{\rm c.c.\; part}\;.
\ea
\ee
Substituting the  fields

\be
\ba{l}
r^0(a_1,b_1)=r^0{}^{\;\alpha(s_1+s_2-1),}{}_{\beta(s_1-s_2-1)}\,a_{1\,\alpha(s_1+s_2-1)}b_1^{\beta(s_1-s_2-1)}\;,
\\
\\
\bar{\omega}^0(a_2,b_2)=\bar{\omega}^0{}_{\;\gamma(s_1+s_2-1),}{}^{\rho(s_1-s_2-1)}\,a_{2\,\rho(s_1-s_2-1)}b_2^{\gamma(s_1+s_2-1)}

\ea
\ee
in  variation (\ref{var_1}) and  using  their Young symmetry properties

\be
\ba{l}
Y_1\equiv S^-_1\;:\quad Y_1\; r^0(a_1,b_1) =0\;,
\\
\\
Y_2\equiv S^+_2\;:\quad Y_2\; \bar{\omega}^0(a_2,b_2) =0\;,
\ea
\ee
we obtain equations of motion that can be  conveniently  written as

\be
\hat{E}\wedge r^0(a,b)=0\;,
\ee
where  $\hat{E}$ is a 2-form differential operator given by

\be
\hat{E}=H^\alpha{}_\beta\,\Big(a_\alpha\frac{\d}{\d a_\beta}+\kappa_2\, b_\alpha\frac{\d}{\d b_\beta}
+\kappa_3\, S^+ a_\alpha \frac{\d}{\d b_\beta}+\kappa_4\, T^+ \frac{\d^2}{\d a^\alpha \d b_\beta}
+\kappa_5 T^+S^+\frac{\d^2}{\d b^\alpha \d b_\beta}\Big)\;
\ee
with the coefficients

\be
\kappa_2=\frac{1+(s_1+s_2-1)(s_2+1)}{1-(s_1-s_2+1)(s_2+1)}\;,\;\;\kappa_3=-\kappa_4=\frac{1-\kappa_2}{2(s_2+1)}\;,\;\;  \kappa_5= \frac{\kappa_2-1}{4s_1(s_2+1)}\;.
\ee
 Analogous equations hold for the complex-conjugated physical field $\bar{\omega}^0$. The operator $\hat{E}$ satisfies the conditions

\be
[S^-,\hat{E}]=0\;,\quad [T^-,\hat{E}]=0\;,
\ee
i.e.,  preserves the  Young symmetry  and  $V$-transversality properties of the physical curvature $r^0$. By construction, this operator also  satisfies the extra field decoupling condition, which means that the term
${\cal T}^-\omega^{1}$ containing the extra field $\omega^1$ in the curvature
$r^0={\cal D}\omega^0+{\cal T}^0\omega^{0}+{\cal T}^-\omega^{1}$ does not contribute to the
equations of motion,i.e., $\hat{E}\wedge {\cal T}^-\omega^{1}=0$.

As in the papers on totally symmetric fields \cite{LV,vf}, we assume that the constraints for extra fields have the form
\be
\label{constr}
\Upsilon_2^+ \wedge r^t(a,b)=0\;,\quad 0\leq t<s_1-s_2-1\;,
\ee
where $\Upsilon_2^+ $ is the 2-form operator that increases $t$ and satisfies the condition
\be
\label{prop}
{\cal T}^+\wedge \Upsilon_2^+ =0.
\ee
The operator $\Upsilon_2^+ $ is required to have property (\ref{prop}) because it ensures
that the number of independent algebraic relations imposed on the curvature $r^t$ coincides with the number of
components of extra fields $\omega^{t>0}$ modulo pure gauge components of the form
$\delta\omega^{t+1} = {\cal T}^-\epsilon^{t+2}$. It  can be shown that the operator $\Upsilon_2^+$ is uniquely fixed in the form
\be
\Upsilon_2^+ = {\cal T}^0\wedge {\cal T}^+\;.
\ee
By virtue of constraints
(\ref{constr}), the field $\omega^{t+1}$ can be expressed via derivatives of $\omega^{t}$ for any $t>0$.
Finally, we can obtain the fields $\omega^{t}$ expressed in terms of the derivatives of $\omega^0$ with the  highest derivative order equal to $t$.

\section{Conclusion}

We have constructed a manifestly covariant Lagrangian formulation for $\ads$ mixed-symmetry
massless gauge fields in the framework of the  $su(2,2)$ spinor formalism. The approach we used is based on the frame-like
formulation of mixed-symmetry fields elaborated in \cite{ASV,ASV2}.
Our results can be regarded as the final step in the study of the manifestly covariant Lagrangian
formulation of $\ads$ higher-spin gauge fields in $su(2,2)$ formalism. An important problem for
further researches is to develop the unfolded form of free mixed-symmetry field dynamics based
on the Weyl tensors following from the equations of motion and constraints for extra fields analyzed
in Sec. 6. This will allow  formulating the central on-mass-shell theorem similarly to the case of totally symmetric gauge fields
\cite{LV,vf,VD5} and establishing a relation  with the unfolded formulation of mixed-symmetry fields developed in \cite{SS2}.
Also, the constructed Lagrangian formulation allows studying  ${\cal N}$ -extended
supersymmetric cubic interactions of $\ads$ gauge fields at the level of action functionals, thus generalizing
${\cal N}=0,1$ results in \cite{VD5,AV1}.

\vspace{1.5cm}

{\bf Acknowledgements}
\\
\\
I am grateful  to M.A. Vasiliev for the useful discussions.
This work is supported in part by the grants RFBR 02-02-16944, LSS 1578-2003-2,
and INTAS 03-51-6346.

\end{document}